\documentclass[10pt]{article}
\usepackage{graphicx,pstricks}
\usepackage{moreverb}
\usepackage{subfigure}
\usepackage{epsfig}
\usepackage{hangcaption}
\usepackage{txfonts}
\usepackage{epsfig}

\title{Generation of pulsed and continuous-wave  squeezed light with $^{87}$Rb  vapor}

\author{Imad H. Agha, Ga\'{e}tan Messin, and Philippe Grangier\\
{\it \small Laboratoire Charles Fabry, Institut d'Optique, CNRS, Universit\'{e} Paris-Sud},\\[-0.8ex]
{\it \small Campus Polytechnique, RD 128, 91127 Palaiseau cedex, France} } 
\date{}
\begin{document}
\maketitle

\begin{abstract}
We present experimental studies on the generation of  pulsed and continuous-wave squeezed vacuum via nonlinear rotation of the polarization ellipse in a  $^{87}$Rb vapor. Squeezing is observed for a wide range of input powers and pump detunings on the D1 line, while only excess noise is present on the D2 line. The maximum continuous-wave squeezing observed is -$ 1.4  \pm0.1$ dB (-2.0 dB corrected for losses). We measure -1.1 dB squeezing  at the  resonance frequency of  the $^{85}$Rb $F=3 \rightarrow F'$ transition, which may allow the storage of squeezed light generated by $^{87}$Rb in a $^{85}$Rb quantum memory.  Using a pulsed pump, pulsed squeezed light with  -1 dB of squeezing for 200 ns pulse widths is observed at 1~MHz repetition rate. 
\end{abstract}

\section{Introduction}
The generation of squeezed states of light has been the subject of intense research since the mid 80's, starting with the pioneering experiment by Slusher \textit{et al.} \cite{slusher85}. Today, there is a strong interest in squeezed light for precision measurement and in particular for gravitational interferometers \cite{grangier87},  as well as for improving the signal-to-noise ratio in communication systems  \cite{saleh}. Additionally, there is a renewed interest in squeezed light with regards to applications in the field of continuous quantum variables \cite{cerf}. 

\medskip Nonclassical states of light are one of the cornerstones of atom-based quantum memories \cite{honda,appel}, as preserving the quantum nature of the state of light is the signature of such a storage system.  A quantum memory, such as that based on the long-lived coherences of an atomic system, is a vital component for the implementation of a quantum repeater \cite{zoller}, which allows for the extension of the range of quantum communication networks. This leads to the desirability of producing pulsed squeezed light resonant with rubidium transitions, possessing a bandwidth comparable to the width of the spectral lines. Recently, there have been several implementations of rubidium resonant squeezed light in optical parametric oscillator (OPO) cavities \cite{Lam, predo} as well as cesium-compatible squeezed light \cite{polzik, burks,polzik2}. These experiments require the use of $\chi^{(2)}$ nonlinear crystals in a double-cavity configuration. The first cavity acts as a frequency doubler to generate the pump beam, while the second acts as the sub-threshold OPO generating the squeezed light. This requires the use of two crystals, two cavities and frequency locking of the two cavities to the input laser beam. In addition, these OPOs do not produce the pulsed squeezing that is ready to be stored in and retrieved from a quantum memory. 

\medskip The generation of resonant squeezed light via nonlinear mixing of a linearly polarized input pump laser with the orthogonal vacuum in a rubidium vapor cell offers an interesting and straightforward alternative to $\chi^{(2)}$-based parametric oscillators. Moreover, using a single-pass configuration makes it possible to use a pulsed-pump input, allowing for the generation of pulsed squeezed vacuum at high repetition rates. This opens the door to experiments involving the de-Gaussification of the generated pulses \cite{alexei}, and  their storage in and retrieval from atom-based quantum memories. Until today, squeezed vacuum generated in a rubidium vapor has been limited to under -1 dB \cite{lvovsky,novikova} of squeezing due to spontaneous emission noise. In this letter, we employ high input pump powers and large detunings from the D1 line of $^{87}$Rb to generate continuous-wave broadband squeezed vacuum (extending from sub-MHz to 20 MHz) with squeezing levels exceeding -1.4 dB  (-2 dB after correction for losses). At the proper pump power and detuning from the $F=2 \rightarrow F'=2$ transition, squeezing is achieved on-resonance with the $F=3 \rightarrow F'$ transition of  $^{85}$Rb, which opens the path for future quantum memory experiments that require resonant nonclassical states of light. As a proof-of-principle, we employ a pulsed-pump input to recover pulsed-squeezed vacuum at MHz repetition rates (1 MHz) and short pulse widths (200 ns).

\medskip The basic mechanism for vacuum squeezing in a single pass linearly polarized pump configuration is polarization self-rotation (PSR) \cite{matsko}, which describes the nonlinear rotation of polarization of elliptically polarized light passing through a medium possessing a $\chi^{(3)}$ Kerr nonlinearity. For a small input ellipticity $\epsilon(0)$, the self-rotation angle (proportional to the nonlinear dispersion) can be written as $\phi=g \epsilon(0) L$, where $L$ is the length of the nonlinear medium, and $g$ is the self-rotation parameter, which for an atomic medium is a function of the input and saturation intensities $I$ and $I_s$, the saturated absorption coefficient, as well as the detuning from the atomic transition. In the quantum regime, and for a linearly polarized pump, the nonlinear ellipse rotation couples the pump to the orthogonally polarized quantum field fluctuations, leading to a phase-sensitive amplification/de-amplification of the vacuum field fluctuations \cite{hsu}, and consequently to anti-squeezing/squeezing effects.

\medskip While nonlinear dispersion decreases with the detuning as $\gamma \Omega^2/ \Delta^3$- where $\gamma$ is the excited-state decay rate, $\Omega$ is the Rabi frequency ($\Omega^2 \propto$ intensity), and $\Delta$ is the detuning from resonance- the linear absorption and hence the spontaneous emission and added quantum noise  decrease as $\gamma^2/\Delta^2$ \cite{hsu} . Consequently, it is desirable to work at large detunings (to reduce absorption) and high input pump powers (to increase non-linear dispersion). In addition to these two experimental parameters, the vapor density inside the cell is another that needs to be optimized in order to maximize the squeezing.  A low vapor density leads to weak nonlinearity, while an extremely high density leads to an increase in spontaneous emission as well as population loss due to collisional relaxation. The interplay of all these parameters leads to an optimal squeezing \cite{lezama,Mikha}.  In our experiment, all three parameters (input power, detuning, and temperature) are varied, in addition to the beam width and the rubidium line (D2 and D1) to find the optimal squeezing parameters. 

\section{Experiment}
The experimental setup is shown in Fig. 1. A Titanium-Sapphire ring laser (Coherent MBR-110) is tuned near the rubidium D1 and D2 lines (795 and 780 nm respectively), with part of the beam sent to a reference cell for calibration. The remainder is spatially filtered with a 50 $\mu$m pinhole and polarized via a Glan-Taylor polarizing prism (GTPP), then focused down into a 7.5 cm rubidium vapor cell, with a  beam waist of 400 $\mu$m. At the output of the cell, the pump and vacuum fields are separated via a polarization beam splitter (PBS). The pump beam is attenuated with a half-wave plate and a prism and used as a local oscillator (LO), while the vacuum port is further purified with a GTPP (providing extinction  of $3\times10^{-5}$ for the pump field). The LO's phase is scanned via a mirror mounted on a piezoelectric transducer (PZT), and recombined with the squeezed/anti-squeezed vacuum port on a 50:50 beam splitter (BS). The outputs of the BS are adjusted with another half-wave/PBS combination to provide true 50:50 splitting and sent via lenses to a balanced homodyne photodetector. A spectrum analyzer records the noise spectral density, whereby the sum of the two photodetectors' (PD1 and PD2) signals provides the laser quantum noise, and the difference the shot-noise limit. The laser is  verified to be shot-noise limited for detection frequencies beyond 1 MHz. Blocking the vacuum port yields the shot-noise level, which is verified to scale linearly with the local oscillator power at all input powers used in the experiment. Next, the cell is heated to provide vapor densities between $10^{10}$ and $10^{12}$ atoms/cm$^3$. The input pump power is varied between 10 and 200 mW, while the detection frequency is varied between 1 and 30 MHz. Near the D2 line, significant excess noise was observed and none of the parameters employed yielded a noise level below the shot-noise limit.

\begin{figure}[b!]
\centering \includegraphics[height= 3 cm]{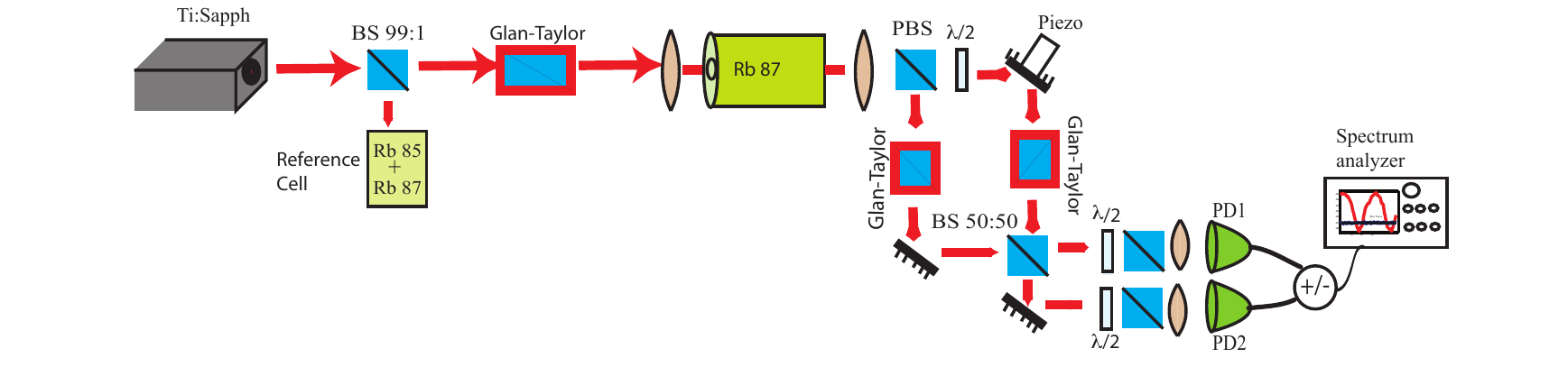}
\caption{Experimental setup: A titanium-sapphire laser is polarization purified and sent to the squeezing cell. The squeezed vacuum is separated from the main pump via a polarization beam splitter (PBS), and the pump, now used as a local oscillator, is attenuated and re-mixed with the squeezed vacuum on a beam splitter and sent into an amplified balanced detector. }
\end{figure}

\medskip One possible explanation for the deviation  of this result from theoretical predictions may be that the models use simplified X-systems \cite{alex} that ignore the hyperfine structure of the excited states. Such an approximation is not expected to be very good for the D2 line which has 4 hyperfine excited  levels, while the D1 line has a simpler hyperfine structure and larger separation between the excited states. It is thus closer to an X-system and hence is more likely  to yield squeezed light via PSR. 

\medskip Fig. 2(a) shows the phase-dependent noise level that is recorded when the pump laser is blue-detuned by 600 MHz from  the $F=2 \rightarrow F'=2$ transition of the D1 line, at an input pump power of 140 mW, cell temperature of 108 $^o$C, and detection frequency of 3 MHz. The observed squeezing is -$ 1.4  \pm0.1$ dB, while the anti-squeezed quadrature is at +$ 5.2 \pm0.1$ dB. This corresponds to -2 dB of squeezing after correction for losses. With the current experimental setup, these figures yielded the highest degree of squeezing. The squeezing as a function of detuning for this value of pump power is shown in Fig. 2(b), where the rapid loss of squeezing for detuning larger than 800 MHz is attributed to the unavoidable presence of  $^{85}$Rb even in the isotopically enriched cell \cite{lett}, which leads to spontaneous emission noise and deterioration of squeezing. This level of squeezing is far below what a simplified theory that does not take into account spontaneous emission predicts \cite{matsko} ($\approx$ 5 dB at our pump power), but interestingly is slightly higher than what is  predicted in the more complete numerical investigation in \cite{lezama}, although that work did not include the entire  hyperfine structure  of the D1 line which could explain the discrepancy. 
\begin{figure}[htbp]
\centering \includegraphics[height=4 cm]{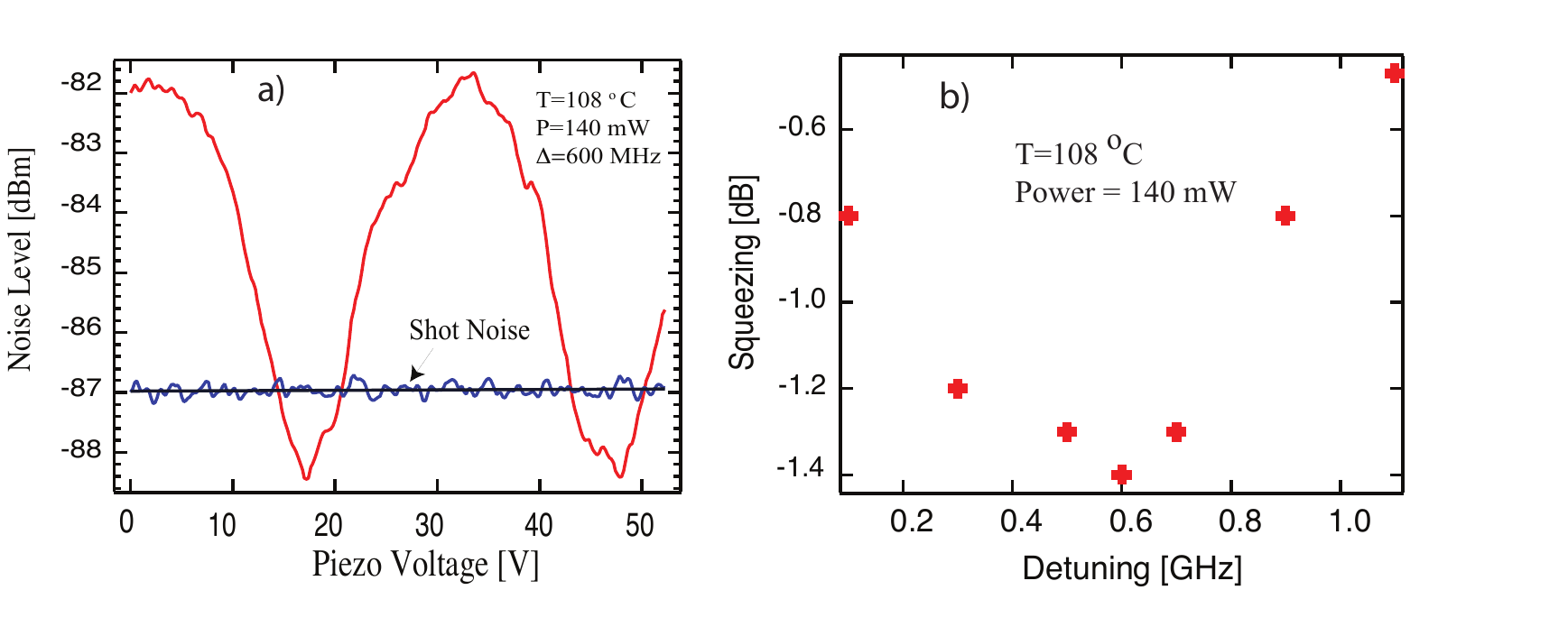}
\caption{a) Output noise level as a function local oscillator phase (piezoelectric transducer voltage). The squeezed quadrature is at -1.4 dB below the shot noise while the anti-squeezed quadrature is at +6.0 dB. b) Squeezing as a function of blue-detuning from the  $F=2 \rightarrow F'=2$ transition at P= 140 mW.}
\end{figure}

\medskip The generation of squeezed light on the D1 line and its absence on the D2 implies that the simpler hyperfine structure, whereby the two excited state transitions act independently, leads to a system that closely resembles the X-system. However, as the pump power is increased, it is expected that the two uncoupled levels, $F=2 \rightarrow F'=2$  and  $F=2 \rightarrow F'=1$ become coupled as the power broadening starts to wash out the hyperfine structure. To verify this point, we run the experiment at different pump powers:  20, 60, 80 and 120 mW. At the lower pump powers, the two energy levels act as independent systems and the squeezing is symmetric on both the  $F=2 \rightarrow F'=2$ and  $F=2 \rightarrow F'=1$ transitions (Fig 3). As the pump power is increased, so does the detuning for optimal squeezing,  which becomes asymmetric on both transitions, indicating a more complicated coupled dynamics than a simple X configuration. At higher pump powers, the squeezing increases until $P\approx 160$mW, after which the squeezing is rapidly lost due to the optimal detuning falling within the doppler profile of the $^{85}$Rb transitions. 
\begin{figure}[tbp]
\centering \includegraphics[width=14 cm, height=4.5 cm]{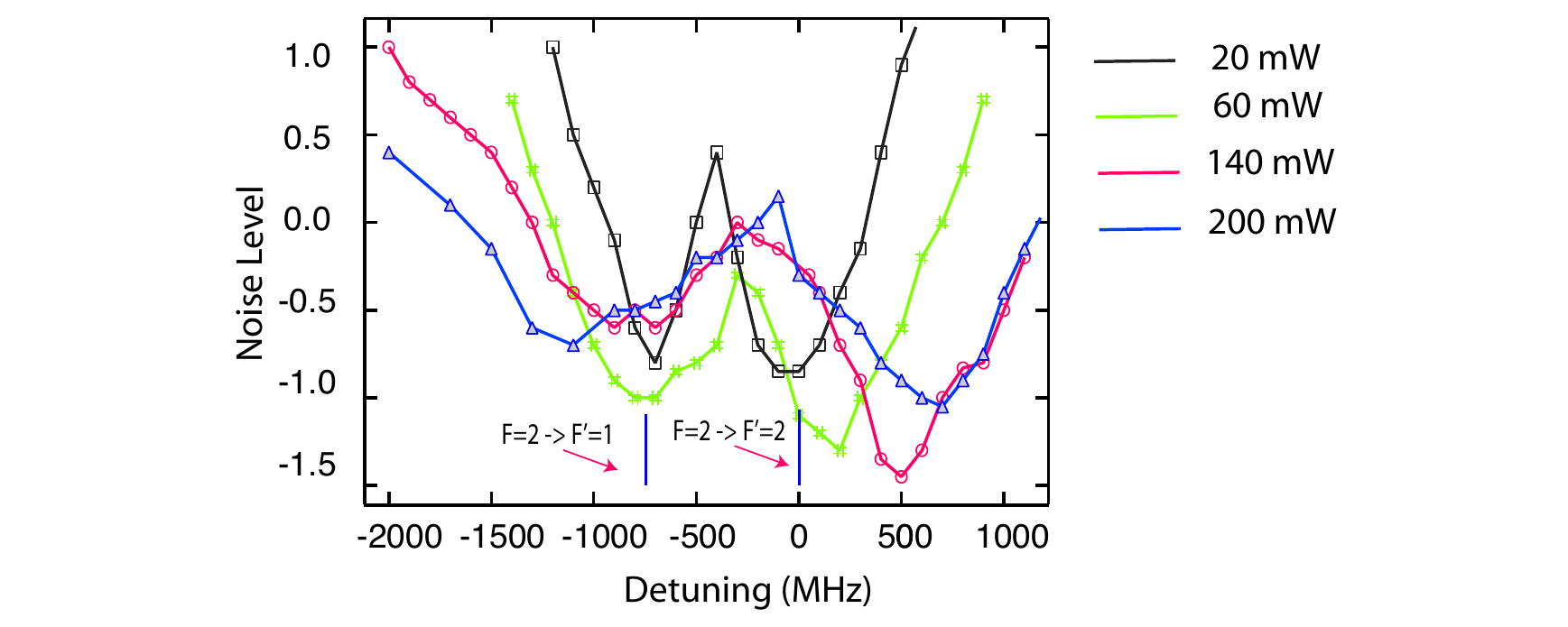}
\caption{Squeezing as a function of detuning for different pump powers. The two vertical lines indicate the $F=2 \rightarrow F'=2$ and the $F=2 \rightarrow F'=1$ transitions.}
\end{figure}

\medskip The interest in generating squeezed vacuum in a rubidium vapor is that it is intrinsically compatible with rubidium atoms with regards to the bandwidth generated. In order to use the generated squeezed state in quantum-memory type experiments, it would be useful to generate the squeezed vacuum on-resonance with rubidium transitions. By blue-detuning our beam from the  $F=2 \rightarrow F'=2$ line, we can generate squeezed vacuum on-resonance with the $^{85}$Rb  $F=3 \rightarrow F'$ transition. The results are shown in Fig. 4(a), where the squeezing is limited to -1.1 dB due to spontaneous emission noise from the excess $^{85}$Rb isotope present in the vapor cell. Figure 4(b) shows the squeezing level as a function of the detection frequency, indicating the presence of broadband squeezing from the low-frequencies (900 kHz) all the way to around 20 MHz. The loss of squeezing at the low frequencies can be attributed to a) The pump laser quantum noise which affects the homodyne detection, b) low-frequency spontaneous emission noise, and  c) phase-mismatch between the pump and the signal due to cross-phase modulation.  These results  open the door to using the current setup in future experiments studying the interaction of resonant squeezed light with a rubidium vapor, whereby the detuning of the squeezed light from the center of the absorption line can be continuously varied between 0 and 700 MHz. The presence of high frequency squeezing renders this scheme compatible with recent proposals and implementations of quantum memories based on Zeeman \cite{ hetet2} and Doppler broadened atomic coherence \cite{gouey} .  
\begin{figure}[htbp]
\centering \includegraphics[width=13 cm, height=4.5cm]{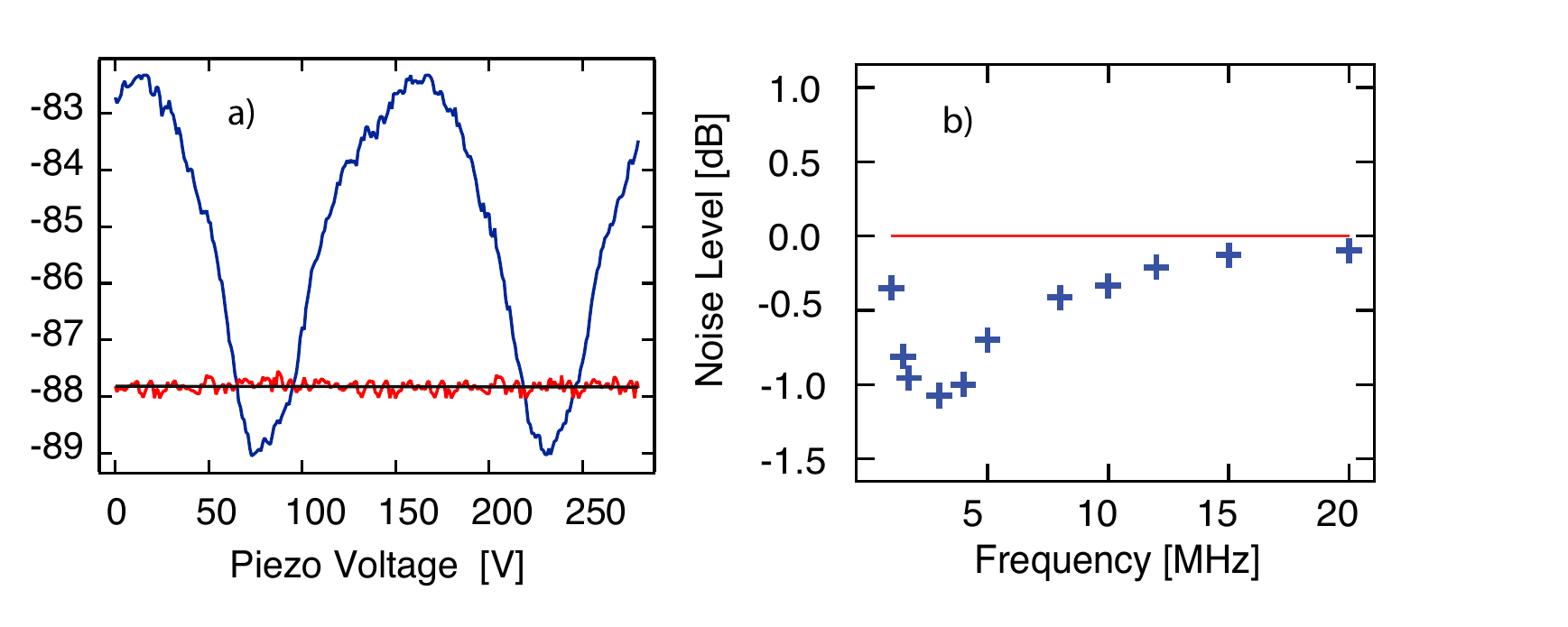}
\caption{a) Noise spectrum with laser tuned on resonance with the $^{85}$Rb $F=3 \rightarrow F'$ transition. b) Squeezing as a function of detection frequency. }
\end{figure} 

\medskip Another advantage for the single pass configuration as opposed to OPO \cite{predo} and cavity-based \cite{josse} configurations   is the ability to use a pulsed-pump input to generate pulsed squeezed light \cite{slusher2} at the output of the vapor cell. This allows for the use of electro-optic and acousto-optic modulators (EOM/AOM)  to chop the pump at the input, or the direct use of rubidium-compatible pulsed lasers \cite{gaetan}, allowing MHz repetition-rate short pulse-width squeezed vacuum generation. A MHz repetition rates imply that single photon detectors can be used in conjunction with homodyne detectors to condition the squeezed vacuum thus producing atom-compatible nonclassical states of light. Such a procedure would not be feasible if the squeezed pulses were to be generated by chopping continuous-wave squeezed vacuum (generated in an OPO)  via a mechanical chopper, since the low repetition rates would lead to a single photon detection probability comparable to detector dark counts. Using EOMs/AOMs to chop continuous-wave squeezed vacuum would degrade the squeezing significantly due to the added quantum noise and losses in these systems. On the other hand, using a single-pass through a nonlinear crystal requires a short pulse laser that is not compatible with narrowband atomic transitions. This renders our source an ideal candidate for producing  nonclassical pulses of light compatible with atomic quantum memories. 

\medskip Since the $\chi^{3}$ nonlinearity is instantaneous and the cross-Kerr squeezing is peak (rather than average) intensity dependent, a pulsed pump should, in theory, produce the same amount of squeezing as a continuous-wave pump, provided the peak powers match. 

\medskip To demonstrate pulsed squeezed vacuum in our single-pass configuration, we use a free-space electro-optic modulator (Linos 0202) to generate 200 ns pulses at 1 MHz repetition rate and 40 mW of peak power, which yields pulses with 6 dB of added quantum noise at our detection frequency (2.7 MHz). Fig 5(a) shows the shot noise over a 5 MHz bandwidth, where the peaks correspond to the repetition rate of the EOM, and the small sidebands to excess noise. The presence of these peaks could be eliminated via balanced homodyne detection with better common-mode rejection which were not available to us at the time.  This technique of detecting pulsed squeezed light, whereby the squeezing is on the sideband of all the pump pulse's spectral components, was theoretically developed by Yurke \textit{et al.} \cite{Yurke}, and subsequently observed by Slusher \textit{et al.} \cite{slusher2} and and Margalit \textit{et al.} \cite{haus}.  By setting the detection frequency between  two of the narrow peaks, we arrive at a shot-noise limited measurement due to a good common-mode rejection ($> 30$ dB).  Fig. 5(b) shows a sample of the phase-dependent noise as a function of the local oscillator phase at 2.7 MHz detection frequency. The squeezing achieved is about -1.0 dB, which is only around 0.2 dB lower than the squeezing obtained at the same peak power in the continuous regime. This method allows for the high repetition-rates required for future experiments involving the conditioning of the squeezed vacuum \cite{alexei} and the generation of non-classical states of light. 
\begin{figure}[tbp]
\centering \includegraphics[width=13 cm, height=4.3cm]{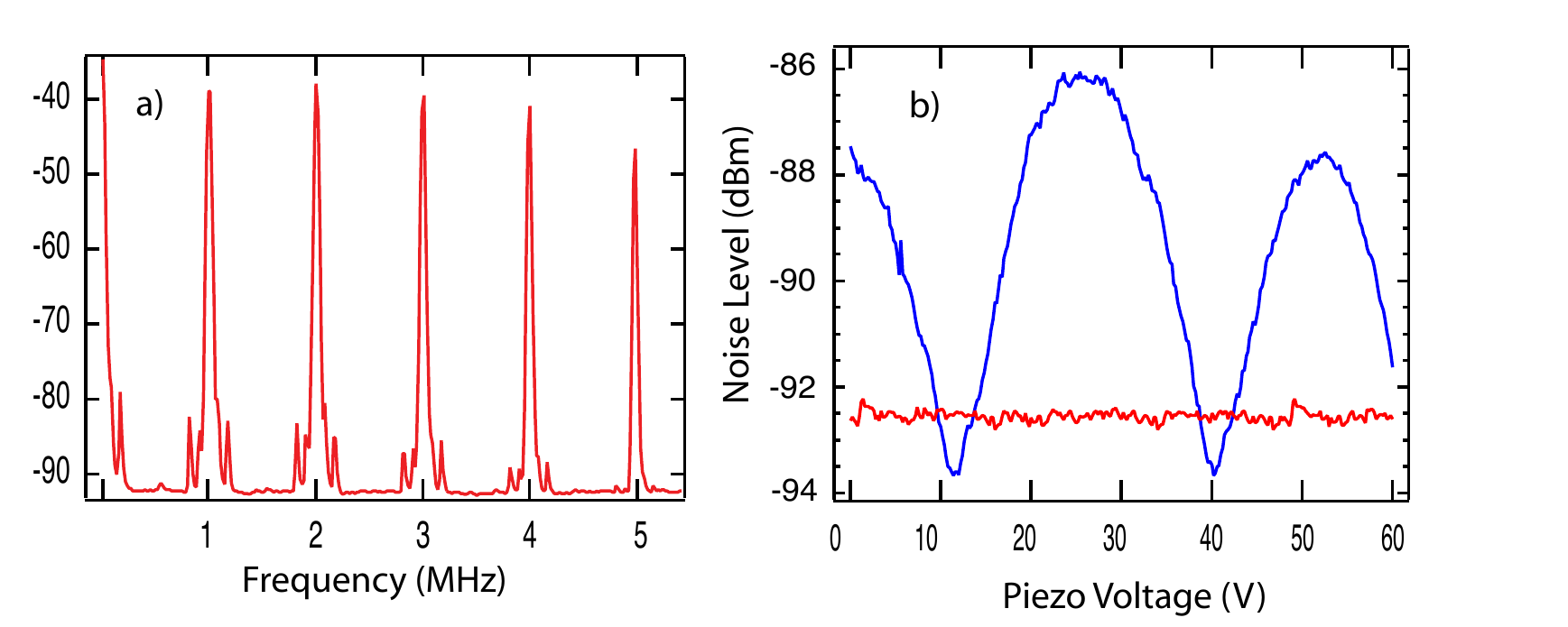}
\caption{a) Output of homodyne detection with squeezed vacuum port blocked showing the (200 ns) pulse spectrum in frequency domain. The detection is shot-noise limited between the modulation peaks. b) Noise level at 2.7 MHz detection frequency showing $\approx$ -1.0 dB of squeezing.}
\end{figure}

\section{Conclusion}
We have demonstrated a simple system based on  nonlinear polarization rotation in a $^{87}$Rb vapor cell capable of producing squeezed vacuum in both the continuous and pulsed regimes under high pump-power conditions with a maximum squeezing exceeding 2.0 dB when corrected for losses. Under suitable pump power conditions, the optimal squeezing produced is on-resonance with $^{85}$Rb transitions. 

\section{Acknowledgements}
This work was supported by the EU grant COMPAS. I. H. A.  gratefully acknowledges the R\'egion Ile-de-France for a postdoctoral grant in the framework of the C'Nano IdF program.   


\end{document}